\documentclass[showpacs,preprintnumbers]{revtex4}
\usepackage{amssymb}
\usepackage{amsmath}
\usepackage{graphicx}
\usepackage{dcolumn}
\usepackage{bm}
\usepackage{epsf}
\usepackage[usenames]{color}

\begin{document}

\preprint{HEP/123-qed}
\title{  Can  realistic nuclear interactions tolerate a resonant tetraneutron? }
\author{Rimantas Lazauskas}
\email{lazauskas@lpsc.in2p3.fr}
\affiliation{DPTA/Service de Physique Nucl\'eaire, CEA/DAM Ile de France, BP 12,
           F-91680 Bruy\`eres-le-Ch\^atel, France}
\author{Jaume Carbonell}
\email{carbonell@lpsc.in2p3.fr}
\affiliation{Laboratoire de Physique Subatomique et de Cosmologie, 53, avenue des
Martyrs, 38026 Grenoble Cedex, France.}
\homepage{http://isnwww.in2p3.fr/theo/Lazauskas/Eng/home.htm}
\homepage{http://isnwww.in2p3.fr/theo/Carbonell/Jaume.html}
\date{\today }

\begin{abstract}
The possible existence of four-neutron resonances close to the
physical energy region is explored. Faddeev-Yakubovsky
equations have been solved in configuration space using realistic
nucleon-nucleon interaction models. Complex Scaling and Analytical Continuation in the
Coupling constant methods were used to follow the resonance pole
trajectories, which emerge out of artificially bound tetraneutron
states. The final pole positions for four-neutron states lie in the third energy
quadrant with negative real energy parts and should thus not be physically
observable.
\end{abstract}

\maketitle

\section{Introduction}\label{sec:Intro}

The existence of pure neutron systems could have far-reaching
implication in nuclear physics \cite{Oglobin}.
However the question
of multineutron existence is far from being  cut-and-dried 
both theoretically and in experiment,  being fed by a long series  
of controversial predictions and measurements.
Recently much attention
have been paid to the possible existence of bound tetraneutron
(four-neutron system). This interest has been triggered by the experimental observation
of  few events in the
$^{14}$Be break-up reaction~\cite{Marques}. On the other hand such
a prospect rises serious objections from the point of view of nuclear
interaction theory. It has been shown by several
groups~\cite{Pieper,Bertulani,Natasha,Thesis} that realistic
nuclear Hamiltonians exclude the existence of bound $^{3}n,$ $^{4}n$
and even larger neutron clusters. In fact, the most favorable
mechanism to construct tetraneutron would be by putting together
two virtual (almost bound) dineutron pairs. However, in order to force
the binding of virtual dineutrons one has to have very strong
neutron-neutron interaction in~\emph{P} and/or higher partial
waves, which is not compatible with our present comprehension of the
nuclear interaction.

Nevertheless the possible existence of resonant states in pure
neutron systems having observable effects in nuclear
reactions, could not be eliminated. Such a scenario is evoked in a recent analysis of
$^{8}He(d,{^{6}Li})4n$ reaction: some excess of low energy $^{6}$Li
nuclei has been observed, which can not be explained by phase space analysis
involving both four free neutrons and two non-correlated dineutron pairs
in the final state~\cite{Exxon}.
Furthermore authors of~\cite{Marques} in their very recent study~\cite{Marques_res}  agreed
that previously observed signals could result from the existence of near-threshold
four-neutron resonance, without involving bound tetraneutron.
The aim of this study is to clarify whether or not the existence of resonant tetraneutrons can be
tolerated by modern nuclear interaction models and thus if these experimental claims can
be supported in a theoretical ground.
This work is a natural extension of our preceding work~\cite{LC_3n_res},
in which we have demonstrated that
realistic nuclear Hamiltonians exclude the existence of physically observable three-neutron resonances.

No proper ab-initio calculations of the resonant tetraneutron with
realistic nn forces are known to the authors.
Some conclusions were drawn in favor of its existence based on calculations
of  tetraneutron bound in an external well~\cite{Pieper}; furthermore it was suggested
that these resonances could have rather large widths.
The only rigorous study of tetraneutron resonances was
accomplished in ref.~\cite{Sofianos} using the simplistic
MT I-III nn interaction, which contains only \emph{S}-waves.
Unfortunately no observable resonances have been found there and
only the existence of some broad subthreshold structures
(S-matrix poles with negative real energy parts) was pointed out.
The same authors remarked however that the positions of these
subthreshold states strongly depend on the details of the  nn
interaction used.
Realistic nucleon-nucleon (NN) models
contain indeed interactions in higher partial waves and are therefore
better suited to accomodate tetraneutron and  push its
resonant states out of the subthreshold region.

\section{Theoretical background}

Although many of the nuclear excited states are resonances,
they are seldom considered in theoretical nuclear structure calculations
due to the huge technical difficulties of solving the continuum states in many-body problems.
These states are often treated as being bound,
but such a procedure is justified only for very narrow
resonances and is not appropriate in our case.
Resonant tetraneutron, if existing at all, will probably have a rather large width.
The problem we are dealing with
represents therefore a double challenge: first it is a four-particle
problem and second, being a continuum state, it has an exponentially diverging wave function.
We will present in what follows the equations allowing to
solve the four-particle problem in a mathematically rigorous way and
describe the methods that make possible the treatment of resonant states.

\begin{figure}[h!]
\begin{center}\mbox{\epsfxsize=12.cm\epsffile{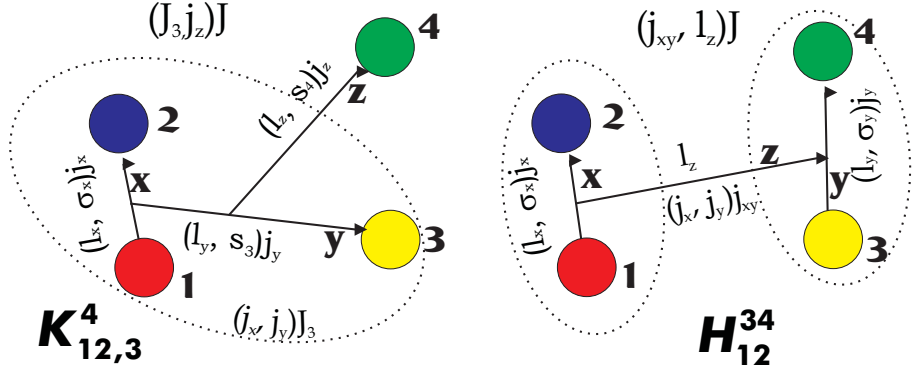}}\end{center}
\caption{Faddeev-Yakubovsky components $K$ and $H$.
Asymptotically, as $z\rightarrow \infty$, components $K$ describe
3+1 particle channels, whereas components $H$ contain asymptotic states of 2+2 channels.} \label{Fig_4b_config}
\end{figure}

In order to solve the four-body problem we decompose the wave
function into a sum of 18 Faddeev-Yakubovsky (FY) components, see
Fig~\ref{Fig_4b_config}, and rewrite the Schr\"{o}dinger equation as
a set of coupled FY equations~\cite{Yakubovski}. If all four particles are
identical, only two of the 18 FY components are independent, which we denote by $K$ and $H$.
These components are furthermore related by two integrodifferential equations:
\begin{eqnarray}
\left( E-H_{0}-V\right) K &=&V(P^{+}+P^{-})\left[ (1+Q)K+H\right]  \notag \\
\left( E-H_{0}-V\right) H &=&V\tilde{P}\left[ (1+Q)K+H\right]  \label{FYE}
\end{eqnarray}%
where $P^{+}$, $P^{-}$, $\tilde{P}$ and $Q$ are particle permutation operators:
\begin{equation}
\begin{tabular}{lll}
$P^{+}=(P^{-})^{-}=P_{23}P_{12};$ & $Q=\varepsilon P_{34}$; & $\tilde{P}%
=P_{13}P_{24}=P_{24}P_{13}$,%
\end{tabular}
\label{PP_operator}
\end{equation}%
and $\varepsilon$ is a Pauli factor related to the exchange of two identical particles:
$\varepsilon =-1$ for two identical fermions.
Using these notations, the four-body wave functions is given by:
\begin{equation}
\Psi =\left[ 1+(1+P^{+}+P^{-})Q\right] (1+P^{+}+P^{-})K+(1+P^{+}+P^{-})(1+%
\tilde{P})H  \label{WF_FY}
\end{equation}

Each FY component $F=(K,H)$ is considered as a function of a
proper Jacobi coordinate set ($\vec{x},\vec{y},\vec{z}$), defined respectively by
\begin{equation}
\begin{array}{ccc}
\vec{x}_{K} & = & \vec{r}_{2}-\vec{r}_{1}\cr\vec{y}_{K} & = & \sqrt{\frac{4}{%
3}}\left( \vec{r}_{3}-{\frac{\vec{r}_{1}+\vec{r}_{2}}{2}}\right) \cr\vec{z}%
_{K} & = & \sqrt{\frac{3}{2}}\left( \vec{r}_{4}-{\frac{\vec{r}_{1}+\vec{r}%
_{2}+\vec{r}_{3}}{3}}\right)%
\end{array}%
\qquad
\begin{array}{ccc}
\vec{x}_{H} & = & \vec{r}_{2}-\vec{r}_{1}\cr\vec{y}_{H} & = & \vec{r}_{4}-%
\vec{r}_{3}\cr\vec{z}_{H} & = & \sqrt{2}\left( {\frac{\vec{r}_{3}+\vec{r}_{4}%
}{2}}-{\frac{\vec{r}_{1}+\vec{r}_{2}}{2}}\right)%
\end{array}%
\end{equation}%
The angular, spin and isospin dependence of these components is expanded using tripolar
harmonics $\mathcal{Y}_{\alpha }(\hat{x},\hat{y},\hat{z})$, i.e:
\begin{equation}
\langle \vec{x}\vec{y}\vec{z}|F\rangle =\sum_{\alpha }\;{\frac{F_{\alpha
}(x,y,z)}{xyz}}\;\mathcal{Y}_{\alpha }(\hat{x},\hat{y},\hat{z}).  \label{KPW}
\end{equation}%
The quantities $F_{\alpha }(x,y,z)$ are called regularized FY amplitudes, and
the label $\alpha $ holds for the set of 10 intermediate
quantum numbers describing a $\left( J^{\pi },T=2,T_{z}=2\right) $ state.
When describing  tetraneutron, the isospin dependence of the FY amplitudes can be
omitted, and the set of quantum numbers $\alpha$ reduces to 8 elements.
We use the $j-j$ scheme for the intermediate coupling of FY amplitudes,  defined as
\begin{eqnarray}
K &\equiv &\left\{ \left[ \left( l_{x}(s_{1}s_{2})_{\sigma _{x}}\right)
_{j_{x}}(l_{y}s_{3})_{j_{y}}\right] _{J_{3}}(l_{z}s_{4})_{j_{z}}\right\} _{J^{\pi }}
\label{PWD} \\
H &\equiv &\left\{ \left[ \left( l_{x}(s_{1}s_{2})_{\sigma _{x}}\right)
_{j_{x}}\left( l_{y}(s_{3}s_{4})_{\sigma _{y}}\right) _{j_{y}}\right]
_{j_{xy}}l_{z}\right\} _{J^{\pi }}
\end{eqnarray}%
where $s_{i}$ is the spin of the individual particle and $J^{\pi
}$ the total angular momentum of the four-particle system. Each of
the $N_{c}=N_{K}+N_{H}$ amplitudes in the expansion (\ref{KPW})
is further conditioned by the antisymmetry properties
$(-)^{\sigma _{x}+l_{x}+1}=\varepsilon $ for $K$ and $(-)^{\sigma
_{x}+l_{x}+1}=(-)^{\sigma _{y}+l_{y}+1}=\varepsilon $ for $H$. FY
components $K$ and $H$ are regular at the origin, and it can
be shown that for bound state problem, they decrease exponentially
outside the interaction domain. In this case one can impose these
functions to vanish on the borders of some constrained box:
\begin{eqnarray}
 F_{\alpha}(x=x_{\max},y=y_{\max},z=z_{\max }) &=&0  \label{BS_BC}
\end{eqnarray}
Equations (\ref{FYE}-\ref{BS_BC}) are enough to solve the bound state problem.

\bigskip

Resonance wave functions are however divergent and cannot be
described by the boundary conditions (\ref{BS_BC}).
In order to solve the resonance problem, we make use of two different
methods, successfully applied in~\cite{LC_3n_res} to treat the
three-neutron system. The implementation of these techniques in
the four-body FY equations is analogous to the three-body case. Therefore
we only briefly discuss them here and  the interested reader can refer to~\cite{LC_3n_res}
for technical aspects.

\bigskip

The method of Analytical Continuation in the Coupling Constant
(ACCC), proposed by Kukulin et al.~\cite{Kukulin}, is based on the fact that
a resonant state arises from a bound one when the interaction between
the particles is made less attractive. The corresponding eigenenergy is
considered as an analytical function of a coupling constant $\lambda $, which
determines the strength of the attractive part of the potential.
Therefore, one can try to analytically continue the energy of the
bound state as a function of the strength $\lambda $ to the
complex plane and obtain this way the width and the position of the resonance.
It can be shown moreover that close to the threshold, where bound
state turns into the resonance, the momenta $k=\sqrt{E-E_{0}}$ is proportional to
\begin{equation}
k\sim x\equiv \left\{
\begin{array}{c}
\lambda -\lambda _{0} \\
\sqrt{\lambda -\lambda _{0}}%
\end{array}%
\right.
\begin{array}{c}
\text{for virtual state} \\
\text{for resonant state}%
\end{array}\label{ACCC_ff}
\end{equation}%
where $\lambda _{0}$ is a critical value of the coupling constant and $%
E_{0}=E(\lambda _{0})$ is the threshold energy. If multiparticle system does not
possess bound states in its subsystems, as it is a case for multineutron,
then $E_{0}=E(\lambda _{0})=0.$

It turns out that using an analytical continuation of $k(x)$ in terms of a
simple polynomial expansion converges slowly and that Pad\'{e} expansion
of order [N,M]
\begin{equation}
k^{n,m}(x)=\frac{a_{1}x+a_{2}x^{2}+...+a_{N}x^{N}}{%
1+b_{1}x+b_{2}x^{2}+...+b_{M}x^{M}}  \label{Pade}
\end{equation}%
is more appropriate.

It is quite simple to put ACCC method in practice. One should
artificially bind tetraneutron by adding some attractive
interaction to the system's Hamiltonian $H=H_{0}+\lambda V_{att}.$
Then, the critical value of the coupling constant ($\lambda _{0}$) 
is determined and  several eigenenergies values $E\left( \lambda
_{i}\right)$ are calculated for $\lambda_{i} >\lambda _{0}$; they
are used to fix the Pad\'{e} expansion~(\ref{Pade})
coefficients. However, to make this
extrapolation efficient, one should provide rather accurate binding energies 
values $E\left(\lambda _{i}\right)$ and an especially precise $\lambda _{0}$ as an input.
While  only few lowest order terms in Pad\'{e} expansion are
enough to determine the positions of narrow nearthreshold
resonances, the description of deep resonances requires several terms
and very accurate input of $E\left( \lambda _{i}\right)$.
The determination of high order Pad\'{e} expansion terms requires at least five digit
accuracy in the binding energies.

\bigskip
The other method we use, namely Complex Scaling
(CS)~\cite{Moiseyev}, can be applied to calculate resonance
positions directly. This method makes use of the similarity transform
\begin{equation}
\widehat{S}=e^{i\theta r\frac{\partial }{\partial r}},  \label{CST}
\end{equation}%
applied to the Hamiltonian of the system, i.e:
\begin{equation}
\left( \widehat{S}\widehat{H}\widehat{S}^{-1}\right) \left( \widehat{S}\Psi
_{res}\right) =E_{res}\left( \widehat{S}\Psi _{res}\right) .
\end{equation}%
Such transformation does not affect the eigenvalue $\left(
E_{res}\right) $ spectra. However, if the scaling angle is large enough 
-- $\theta >\frac{1%
}{2}\left\vert \arg E_{res}\right\vert $ -- the modified resonance eigenfunctions $%
\left( \widehat{S}\Psi _{res}\right) $ become square integrable. 
Evidently, the
CS method can be applied to FY equation. By this transformation
all the radial variables $r\equiv (x,y,z)$ in eq.(\ref%
{FYE}) are replaced by $re^{i\theta }\equiv (xe^{i\theta
},ye^{i\theta },ze^{i\theta })$. The problem becomes analogous 
to a bound state one with complex variable and
transformed FY amplitudes $\left( \widehat{S}F\right)$, which
unlike resonance eigenfunctions $\Psi _{res}$ are in Hilbert space.

CS transformation requires the analytical continuation of the potential 
$V(x)$ into the complex plane  $V(xe^{i\theta })$. This turns out to
be a weak point of this method when applied to nuclear systems,
since as discussed in~\cite{LC_3n_res}, nuclear potentials have
mischievous analytical properties: they become strongly
oscillating and even divergent already for relatively small
transformation angles $\theta >30^{\circ }$. This fact
limits the applicability of CS method to narrow resonances, with 
$Im(-E_{res})<2Re(E_{res})$ values.

The numerical solution of FY equations is performed by expanding
$F_{\alpha }(xyz)$ on a basis of three-dimensional piecewise
Hermite polynomials and projecting equation (\ref{FYE}) with boundary
conditions (\ref{BS_BC}) onto  tripolar harmonics. In this
way, the integro-differential FY equations are converted into a linear
algebra problem:
\begin{equation*}
AX=E_{res}BX
\end{equation*}%
with $A$ and $B$ being large square matrices, whereas
$E_{res}$ and $X$ are respectively the eigenvalue and the eigenvectors  to be determined. 
The reader interested in a detailed
discussion on the formalism and the numerical methods used should refer to~\cite{Thesis}.

\section{Results and discussion}

The results presented below have been obtained by using the charge-symetry breaking 
Reid 93 potential to describe the nn interaction.
This choice is dictated by purely practical reasons: as discussed
in~\cite{3n_res_gloe,LC_3n_res}, Reid 93 model has better
analytical properties to perform the complex scaling operation
(\ref{CST}) than his coordinate-space modern concurrents.
We would like to remark however that other realistic NN interaction
-- namely AV14, AV18 and Nijm II -- provide very similar results for
two- and three-neutron systems and exhibit also a similar behavior
for the -- artificially bound -- tetraneutron. These facts let
us believe that no qualitative changes in the four-neutron resonance
can emerge from the properties of a  particular model.
All the calculations presented in what follows use the value
$\frac{\hbar ^{2}}{m_n}=41.44$ MeV$\cdot $fm$^{2}$ as an input for the neutron mass.

\bigskip
In a similar way as in the study of three-neutron system \cite{LC_3n_res},
we introduce an additional attractive four-nucleon (4N) force to analyze
the tetraneutron resonance trajectories in a systematic way. We have chosen the form:
\begin{equation}\label{Yukawa_3NF}
V_{4n}=-W \; \rho \; e^{-{\rho\over\rho_0}},
\end{equation}
where $W$ and $\rho_0$ are respectively  the strength and range parameters of
the potential,  and the hyperadius $\rho =\sqrt{x^{2}+y^{2}+z^{2}}$ 
is an invariant quantity with respect to the permutation operators (\ref{PP_operator}).  
Such kind of force is easy to implement in  FY equations  (\ref{FYE}).
In our previous work \cite{LC_3n_res} devoted to the three-neutron system,
we have used a 3n force having the standard Yukawa form.
However we have found the functional form (\ref{Yukawa_3NF}) more appropriate for
studying artificially bound tetraneutron. This form does not
diverge as $\rho\rightarrow0$ and thus avoids a rapid shrinking of the bound structures
generated.

As it has been already remarked in~\cite{Thesis,Natasha,Pieper}, 
an extremely strong additional interaction is required
to force tetraneutron binding. As a consequence, the thus generated  bound system
is a very compact object making unlikely its physical existence. 
On the other hand,
resonances are extended structures. In order to ease the transition
from bound to resonant tetraneutron, we have fixed a rather large
value for the range  parameter $\rho_0$ in (\ref{Yukawa_3NF}) and taken $\rho_{0}=2.5$ fm,
a value considerably larger than the one we could expect for  a realistic 4N interaction.

Our strategy to study 4n-resonances is to vary the strength of the potential
$W$ and trace the resonance energy-trajectory $E_{res}(W)$.
The final
resonance positions, which correspond to realistic nuclear interaction, 
are eventually reached at $E_{res}(W=0)$.

When applying ACCC method, the parameter $\lambda$ given in (\ref{ACCC_ff}) is identified
to the 4N force strength $\lambda\equiv W$. We determine several
auxiliary values of $E_{i}(\lambda _i) $ in the bound tetraneutron
region $\lambda _i>\lambda _0$. These values are later used as an
input to determine the Pad\'{e} expansion coefficients $\left(
a_{i},b_{j}\right) $ of eq.(\ref{Pade}). Few
calculations are performed in low energy region as well to determine the
critical value $W _0 \equiv\lambda _{0}$ for which tetraneutron is
bound with zero energy $E_{res}(W _{0})=0$.
Once $\left( a_{i},b_{j}\right) $ and $\lambda _{0}$ are known, we use equation 
(\ref{Pade}) to  analytically continue $E_{res}(W)$ curve to the resonance region $W<W _{0}$.

When applying CS method, we perform a series of direct resonance calculations
for several decreasing values of $W<W_{0}$, until the calculations
become unstable due both to the large size of the resonance widths and to the necessity of
using ever increasing scaling parameter $\theta $. 
This method is used in the near threshold region --
$W\lessapprox W _{0}$ and Im$(-E_{res})<<$Re$(E_{res})$  --
where it gives very good results to improve the
accuracy of $\lambda_{0}$ used in ACCC calculations.

\begin{table}[h!]
\begin{tabular}{ccccccc}\hline
$J^{\pi}$ & $0^{-}$ & $1^{-}$ & $2^{-}$ & $0^{+}$ & $1^{+}$ & $2^{+}$ \\
\hline $W_0 $ & 38.70 &  38.67 &  38.68   & 22.90 & 22.92 &  40.38\\
       $W'$   &  3.0 &  3.2 &  3.9   & 3.5 & 3.6 &  4.1\\
       $E_{res}(W=0)$  &-1.0-9.9i  & -1.1-9.8i &-1.4-9.7i   &-1.1-6.3i &-1.1 -6.5i & -1.4-10.9i   \\\hline
\end{tabular}

\caption{
Critical strengths $W _{0}$ (MeV$\cdot$fm$^{-1}$) of the phenomenological  force 
(\protect\ref{Yukawa_3NF}) required to bind four neutrons in
different states. The range parameter $\rho_{0}$ of this force was fixed to 2.5
fm. $W'$ are the strength values  at which
tetraneutron becomes subthreshold, i.e. Re$(E_{res})=0$. In the last row of the table
are given the physical resonances positions ($W=0$).}\label{Tab_val_w0}
\end{table}
\bigskip

In  Table~\ref{Tab_val_w0} are  summarized (second row)  the critical
strengths $W _0$ required  to bind tetraneutron in states with different $J^{\pi}$ quantum numbers.
Even though we have taken 4N force with a rather long range,
the critical strengths values $W _0$ are still considerable.
Other noticeable feature is that these critical values are almost equal for
all negative parity tetraneutron states
$J^{\pi }=0^{-}$, $1^{-}$ and $2^{-}$ we have considered. 
The reason for such a degeneracy is that tetraneutron binding energies are insensitive to nn
interaction in \emph{P} and higher partial waves. Actually, their values remain unchanged up to
three-four digits if these $l_{nn}\geq1$ interaction terms are switched off.
Tensor coupling is present only in $l_{nn}\geq1$ partial waves  
and has very small impact on these states. As a consequence the total spin (S=1) and
angular momenta (L=1) are separately conserved.

A similar situation is observed for positive parity states $J^{\pi}=0^{+}$ and $1^{+}$, 
which are also almost degenerate. These
states are dominated by the FY amplitude of type K with
$l_x=l_y=l_z=0$ intermediate quantum numbers, i.e. they are almost pure L=0 states. 
On the other hand, they differ by their total spin (S=0 for $J^{\pi }=0^{+}$ and
S=1 for $J^{\pi }=1^{+}$). 
Unlike expected, the $J^{\pi }=0^{+}$ state has a structure dominated by 
a "nn+n+n" type configuration  and not by a "nn+nn" one, i.e. containing only
a single $^1S_0$ dineutron pair and not two. 
For $J^{\pi }=2^{+}$, $W _{0}$  is considerably larger. 
This state must have a total spin S=2 to be
realized with L=0, i.e. all neutron spins pointing in the same
direction and thus no any $^1S_0$ dineutron pairs with antiparallel neutron spins
is present. The corresponding large $W _{0}$ value
can therefore be understood as the price to pay for breaking the remaining dineutron pair. 
$J^{\pi }=0^{+}$ and $1^{+}$
states remain also unchanged if nn \emph{P}-wave interaction is switched off. 
A very strong enhancement of these waves is required, as much as creating a dineutron resonance!, 
in order to see their effect in the binding energies. 
The only state sensible to nn \emph{P}-waves is the $2^{+}$.

\begin{figure}[h!]
\begin{center}
\mbox{\epsfxsize=11.cm\epsffile{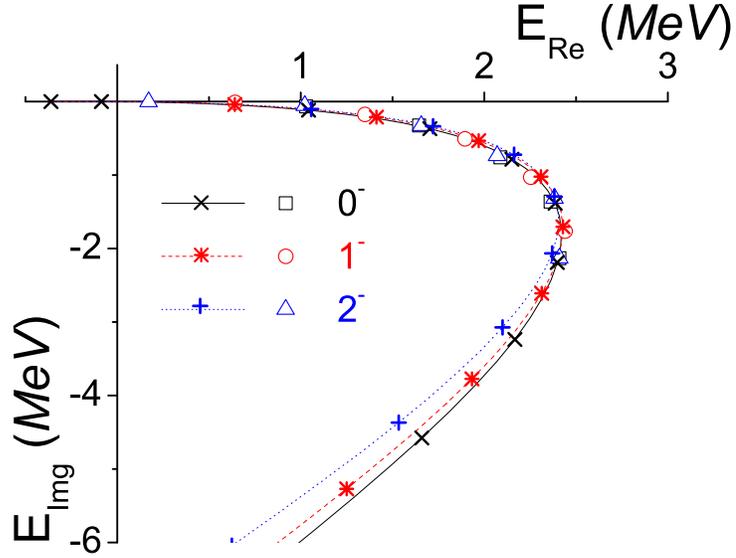}}
\hspace{-1.cm}
\caption{Negative parity tetraneutron resonance trajectories parametrized by the strength
$W$ of the phenomenological 4NF.
ACCC results are denoted by lines with overimposed $x,*,+$ symbols.
They correspond to $W$ values by steps of 4 MeV$\cdot$fm$^{-1}$ 
starting from 38 MeV$\cdot$fm$^{-1}$  for
$J^\pi=0^-$ and $2^-$ states and from 36 MeV$\cdot$fm$^{-1}$ for $1^-$. 
CS results are represented by  circles, squares and triangles. } \label{Fig_01m_traj}
\end{center}
\end{figure}
\begin{figure}[h!]
\begin{center}
\mbox{\epsfxsize=11.cm\epsffile{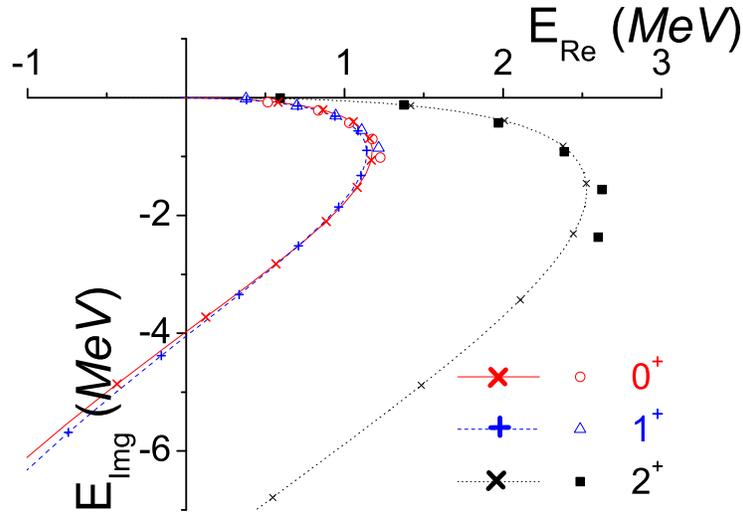}}
\end{center}
\hspace{-1.5cm}
\caption{The same as in figure \ref{Fig_01m_traj} for positive parity states.
$W$ values are reduced from 20
MeV$\cdot$fm$^{-1}$ with step of 2 MeV$\cdot$fm$^{-1}$ for
$J^\pi=0^+$, from 21 MeV$\cdot$fm$^{-1}$ in step of 2
MeV$\cdot$fm$^{-1}$ for $1^+$ and from 38 MeV$\cdot$fm$^{-1}$ in
step of 4 MeV$\cdot$fm$^{-1}$ for $2^+$.} \label{Fig_01p_traj}
\end{figure}

\bigskip
In figures~\ref{Fig_01m_traj} and \ref{Fig_01p_traj}  are displayed
the tetraneutron resonance trajectories for the same negative and
positive parity states we have considered in Table \ref{Tab_val_w0}.  
In both figures, CS results are
indicated using empty symbols (square, cercle and triangle) which correspond
to different values of the 4NF strength parameter W. 
ACCC trajectories for different states are depicted by solid, dashed and dotted lines.
They have overimposed star-like
$x,*,+$ symbols  which correspond to the same W values than those used in CS calculations. 
In order to compare quantitatively the  agreement between both methods,
the numerical values of some resonance positions are also given in Table~\ref{tab_compa}.
This agreement is rather nice for narrow resonances. 
For wider resonances, small discrepancies appear,
which are due to the drawbacks present in CS method described above. 
Sizeable differences appear for resonances with Re$(E_{res})<-$Im$(E_{res})$, which is
the limit of applicability of CS transformation. 

\begin{table}[h!]
\begin{tabular}{ccccccc}\hline
 &\multicolumn{3}{c}{CS} &\multicolumn{3}{c}{ACCC}\\ \cline{2-4}\cline{4-7}
$W $  & $0^{-}$ & $1^{-}$ & $2^{-}$ & $0^{-}$ & $1^{-}$ & $2^{-}$ \\ \hline
30.0     & 1.67  -0.33i  & 1.68  -0.33i &1.67   -0.33i  &1.70  -0.36i &1.72  -0.35i & 1.72  -0.34i \\
24.0     & 2.24  -1.03i  & 2.24  -1.02i & 2.24  -1.02i  &2.30  -1.05i &2.31  -1.03i & 2.30  -1.00i \\
20.0     & 2.41  -1.67i  & 2.41  -1.67i & 2.41  -1.66i  &2.42  -1.75i &2.43  -1.72i & 2.41  -1.67i\\
\hline
 &\multicolumn{3}{c}{CS} &\multicolumn{3}{c}{ACCC}\\\cline{2-4}\cline{4-7}
$W $  & $0^{+}$ & $1^{+}$ & $2^{+}$ &
$0^{+}$ & $1^{+}$ & $2^{+}$ \\ \hline
34.0    &  &  & 1.37 -0.12i&&&  1.41 -0.13i\\
26.0    &  &  & 2.39 -0.92i&&&  2.38 -0.84i\\
18.0    &0.84  -0.22i  &0.84  -0.22i  &2.60 -2.46i    &0.85  -0.21i  &0.84  -0.22i  & 2.44  -2.31i\\
15.0    &1.11  -0.56i  &1.11  -0.55i  &               &1.12  -0.54i  &1.09  -0.56i  & 2.22  -3.12i \\
13.0    &1.22  -0.85i  &1.21  -0.86i  &               &1.17  -0.86i  &1.14  -0.89i  & 1.98  -3.76i  \\\hline
\end{tabular}%

\caption{Comparison of CS and ACCC method results. Resonance positions for 
tetraneutron states obtained by adding phenomenological  4n 
force with strength W (in MeV$\cdot$fm$^{-1}$) and range 
 $\rho_{0}$=2.5 fm are compared.}
\label{tab_compa}%
\end{table}

In this region, ACCC results are still rather well converged (better than $~5$\%) 
with respect to the Pad\'{e} expansion. However this convergence
is getting worst when one departs further and further  from the bound state region.
The accuracy of Pad\'{e} expansion is 20\% near the subthreshold region, where
resonance trajectory moves into the third energy quadrant. 
The accuracy of the physical resonance positions, when the additional interaction is
fully removed ($W=0$), is at worst of 50\%. 

\begin{figure}[h!]
\begin{center}
\mbox{\epsfxsize=14.cm\epsffile{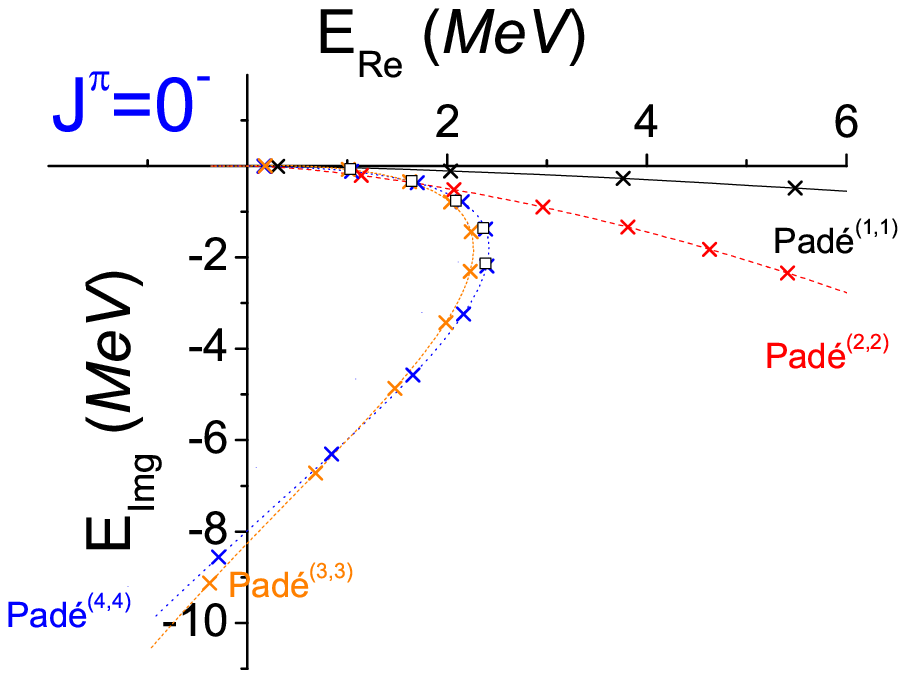}}
\end{center}
\caption{Convergence of ACCC method with respect to the order [M,N] of
Pad\'{e} expansion for $J^{\pi }=0^{-}$ tetraneutron state. 
ACCC curves are followed by star-like points indicating the resonance positions
for W values decrasing from 36 MeV$\cdot$fm$^{-1}$ by step of 4 MeV$\cdot$fm$^{-1}$. 
CS results are 
presented by full cercles and corresponding to W values from 36 to 15 MeV.fm.} \label{Fig_conver}
\end{figure}
These limitations in ACCC accuracy are due to the increasing size of the Pad\'{e} expansion 
argument
$x=\sqrt{\lambda -\lambda _{0}}$, which forces to take into account higher order terms. 
The precise determination of high order Pad\'{e}
coefficients fails due to the severe accuracy criteria it imposes to the input. 
As an example, we have illustrated in figure~\ref{Fig_conver}, the Pad\'{e}
expansion convergence for $J^{\pi }=0^{-}$. 
We can see that the shapes for
[N,M]=[3,3] and [4,4] order Pad\'{e} expansion curves
are already very close to each other. However the separation between the energies
corresponding to the same 
W values still exist and it increases when one departs from the bound state region.

\begin{figure}[h!]
\begin{center}
\mbox{\epsfxsize=14.cm\epsffile{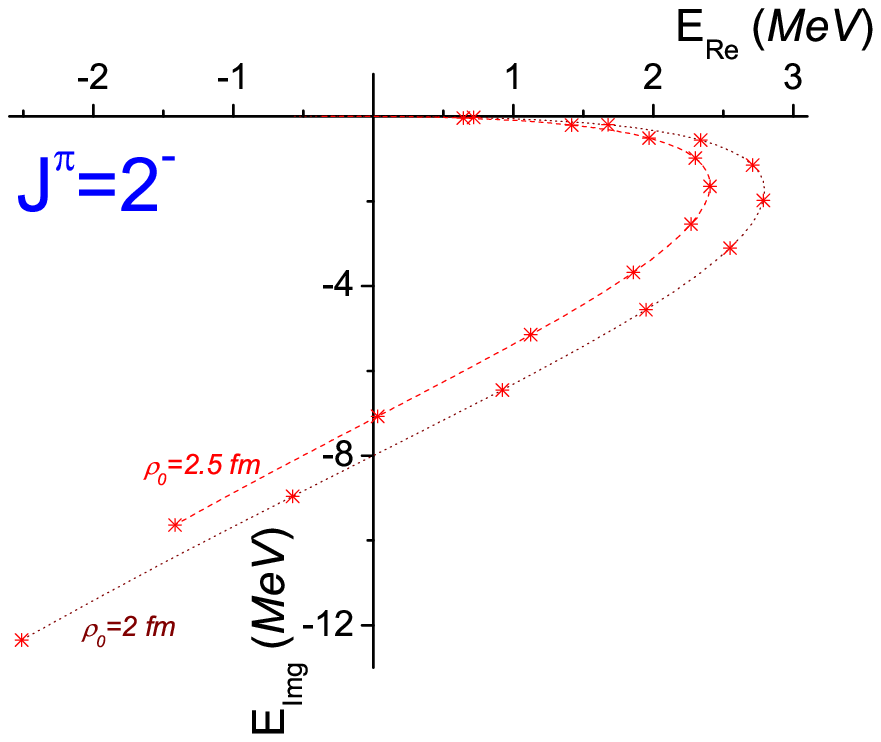}}
\end{center}
\caption{Comparison of resonance trajectories for $J^{\pi }=2^{-}$
tetraneutron, two different curves correspond calculations with
4NF eq.(\ref{Yukawa_3NF}) having length $\rho_{0}$=2.5 (dashed
curve) and 2 fm (dot curve). The points correspond resonance
positions for W being reduced from 36 in step of 4
MeV$\cdot$fm$^{-1}$ for $\rho_{0}$=2.5 fm curve and from 72 in
step of 8 MeV$\cdot$fm$^{-1}$ for $\rho_{0}$=2 fm curve.}\label{Fig_2minus}
\end{figure}

The uncertainty  in determining the final
resonance positions is also manifested in figure~\ref{Fig_2minus}.
In this figure, we compare
the resonance trajectories for $J^{\pi }=2^{-}$ tetraneutron state obtained with 4NF 
of eq.(\ref{Yukawa_3NF})
having different values of the range parameter $\rho_{0}$. For $\rho=2.5$ fm 
we ended with a value $E_{res}=-1.4-9.7i$ MeV,
while for $\rho_{0}=2$ fm  with $E_{res}=-2.5-12.5i$ MeV.

\bigskip
As it has been discussed above, tetraneutron negative parity states on one hand
and $J^{\pi }=0^{+}$ and $1^{+}$ ones on the other hand, are almost degenerate in energy.
This degeneracy is also reflected in the corresponding resonance trajectories, 
which superimpose close to the threshold. 
Notice however that the small difference -- not exceeding several keV -- in the binding energies,
results into an increasing separation of these curves. 
This demonstrates the necessity of performing very accurate inputs for the Pad\'{e} extrapolation
and the difficulty of obtaining accurate predictions for broad resonances in the ACCC method.

Regardless the convergence problems mentioned above, our results
indicate that the final resonance positions will always stay in the third energy
quadrant for all tetraneutron states. 
An accurate determination of the physical resonance position
is not possible with the methods used in the present work.
Nevertheless, for all calculations we have performed, they
were situated in the third energy quadrant (Re(E)$<0$,Im(E)$<$0).
The approximate values for their positions obtained with the ACCC method
and 4NF of equation (\ref{Yukawa_3NF}) are summarized in the
last row of Table~\ref{Tab_val_w0}.
The accuracy of these results is estimated to be of 50\%.

In Table~\ref{Tab_val_w0} we have also displayed the strengths $W'$ of
the 4N force (\ref{Yukawa_3NF}) at which the resonance trajectories
crosses the imaginary-energy axis, slipping from fourth into
third energy quadrant. One can argue that these values are pretty
small and that a small correction of nuclear interaction (like
the presence of attractive three-nucleon force) can push tetraneutron
states back to the fourth energy quadrant (with positive real energy
parts). On this point we would like to mention that the smallness of $W'$ is
only apparent and entirely due to the unrealistic long range character of
the 4NF we have chosen. $W'$  value would increase drastically if
the range of the potential $\rho _0$ is reduced to make 4NF more realistic . 
This fact is demonstrated in figure~\ref{Fig_2minus}, where $W'$ value for $2^{-}$
tetraneutron state increases from 3.9 to 10.8  MeV$\cdot$fm$^{-1}$  
when $\rho_{0}$ is reduced from 2.5 to 2 fm. This result shows
that any realistic ($\rho<1.4$ fm) multineutron force should be very strong to
keep a multineutron resonances in the fourth energy quadrant.

\begin{figure}[h!]
\begin{center}
\mbox{\epsfxsize=14.cm\epsffile{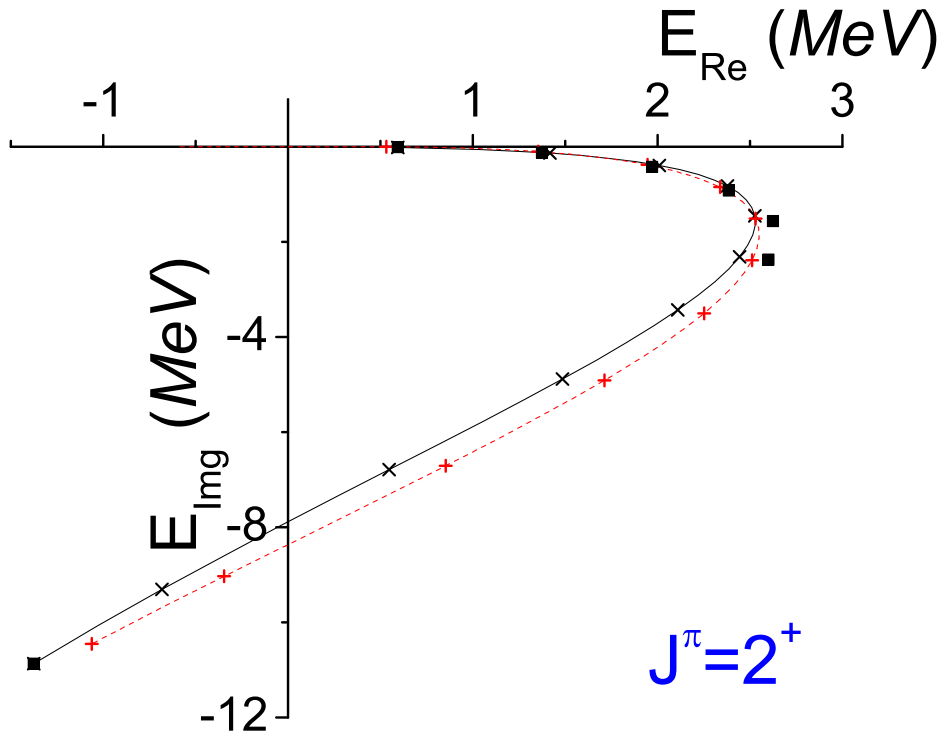}}
\end{center}
\caption{Sensibility of the  $2^{+}$ tetraneutron resonance trajectory
with respect to nn \emph{P}--waves. Solid line correspond to Reid 93 nn
interaction and dashed line was obtained with  nn
\emph{P}--waves enhanced by a factor $\gamma$=1.2} \label{Fig_2plus}
\end{figure}

\bigskip
The $J^{\pi }=2^{+}$ tetraneutron state represents an interesting
case, since it shows the largest sensitivity to nn \emph{P}-waves interaction. 
Some 3N and 4N scattering observables which are difficult to be reproduced with the existing
models indicate a strong nn $\emph{P}$-waves contribution. 
It has  been suggested that these
discrepancies in 3N and 4N scattering observables can be significantly improved by modifying
 nn \emph{P}-waves within 20\%~\cite{P_waves_Gloe,P_waves_Pisa,Contra_PW}. 
We have explored  such a possibility and traced in fig.~\ref{Fig_2plus} the tetraneutron resonance
trajectory for Reid 93 interaction with nn \emph{P}--waves enhanced by a factor $\gamma$=1.2.
By doing so a slightly weaker critical strength of 40.02 MeV$\cdot$fm$^{-1}$ is required 
to bind tetraneutron, in comparison with 40.38 MeV$\cdot$fm$^{-1}$ of the original Reid 93 force. 
However, apart from a small shift in the  $E _{res}(W)$ trajectory, such
a modification of nn \emph{P}--waves has not changed its qualitative
behavior, ending up very close to its original value, always located in the third energy quadrant. 
These nn P-waves should be much more strongly enhanced, as
much as creating dineutron resonances, to
result in sizeable effects in tetraneutron resonance positions.

\bigskip
Finally, we would like to remark that
even if there was a resonance in the fourth energy quadrant having
a small real energy part and a large imaginary one,
it would be difficult to identify it experimentally.
Resonance should have a rather small width $\Gamma=-2 Im(E_{res})$ to produce
a visible effect in the experimental cross section 
and a $E=Re(E_{res})$ centered  Breit-Wigner shape. 
At most, it will give a weak enhancement in the cross section,
hardly discernable from the background and not necessary centered
around the $E=Re(E_{res})$. 
This makes very doubtful the
perspective of physically observable tetraneutron resonances. 
Their eventual existence would imply a too strong modifications in the present nuclear Hamiltonians.

\bigskip
Our results are in qualitative agreement  with the findings of
Sofianos et al.~\cite{Sofianos}, where authors were able to
accurately determine the tetraneutron resonance positions in the third
energy quadrant for positive parity states, although using S-wave
MT I-III potential. Due to the small influence of \emph{P} and higher nn
partial waves on tetraneutron states, \emph{S}-wave models become
very appropriate to study this system.

\section{Conclusion}\label{sec:Conclu}

Configuration space Faddeev-Yakubovsky equations have been solved
with the aim of determining the positions of the four-neutron resonances
in the complex energy plane.

Realistic Reid 93 nn interaction model has been used. A systematic
study of four-neutron resonances have been accomplished by first adding
to the nuclear hamiltonian an attractive four-neutron force to artificially bind tetraneutron.
The trajectory of the energy eigenvalue is then traced as a function
of the strength of the additional force until it is fully removed. 

Two methods, namely  Complex
Scaling and Analytical Continuation in the Coupling Constant, were
employed to follow these trajectories. 

The low lying four-neutron resonance trajectories, 
corresponding to states with quantum numbers $J^{\pi}=0^{\pm},1^{\pm},2^{\pm}$, 
were shown to settle in the third energy quadrant (Re(E)$<0$, Im(E)$<0$))
well before the additional 4n force is completely
removed. Furthermore, these resonances acquired a rather large
imaginary energy $\Gamma=2{\rm Im(-E)}\approx 20$ MeV and 
should hardly be experimentally observable. 
Tetraneutron compound -- bound or resonant --
can be created only in strong external fields and
would disintegrate right after such a field is removed.

Finally, we have demonstrated that the four-neutron physics is
entirely determined by nn \emph{S}-waves, namely $^1S_0$ one, which is
controlled by the experimentally measurable nn-scattering length.
All realistic nuclear interaction models 
should thus provide qualitatively identical results for
tetraneutron resonances. 
This fact is supported by performing 
similar studies binding artificially tetraneutron states using various NN
models~\cite{Thesis}.

\bigskip
{\bf Acknowledgements:}
Numerical calculations were performed at
Institut du D\'eveloppement et des Ressources en Informatique Scientifique (IDRIS) from  CNRS
and at Centre de Calcul Recherche et Technologie (CCRT) from CEA Bruy\`eres le Ch\^atel.
We are grateful to the staff members of these two organizations for their kind hospitality and useful advices.

\end{document}